\documentclass[aps,pra,onecolumn,superscriptaddress]{revtex4}

\usepackage{longtable}
\usepackage{graphicx}
\usepackage{amsmath}
\usepackage{amssymb}
\usepackage{amsthm}

\usepackage{amsfonts,epsfig}
\usepackage{latexsym}
\usepackage{amsfonts}
\usepackage{mathrsfs}
\usepackage{natbib}
\usepackage{color,verbatim,graphics}
\usepackage{psfrag}
\DeclareMathAlphabet{\mathrsfs}{U}{rsfs}{m}{n}
\DeclareMathAlphabet{\mathpzc}{OT1}{pzc}{m}{it}
\DeclareMathAlphabet{\matheus}{U}{eus}{m}{n}
\DeclareMathAlphabet{\mathbbold}{U}{bbold}{m}{n}

\newcommand{\be}{\begin{equation}}
\newcommand{\ee}{\end{equation}}
\newcommand{\ba}{\begin{eqnarray}}
\newcommand{\ea}{\end{eqnarray}}
\newcommand{\ban}{\begin{eqnarray*}}
\newcommand{\ean}{\end{eqnarray*}}

\newcommand{\expect}[1]{\langle #1 \rangle}
\newcommand{\tr}{\mathrm{tr}}

\newcommand{\one}{\leavevmode\hbox{\small1\normalsize\kern-.33em1}}%

\newcommand{\etal}{{\it{et al.}}}

\begin{document}

\title{Information-Causality and Extremal Tripartite Correlations}

\author{Tzyh Haur Yang}
\affiliation{Centre for Quantum Technologies, National University of Singapore, 3 Science drive 2, Singapore 117543}
\author{Daniel Cavalcanti}
\affiliation{Centre for Quantum Technologies, National University of Singapore, 3 Science drive 2, Singapore 117543}%
\author{Mafalda L. Almeida}
\affiliation{Centre for Quantum Technologies, National University of Singapore, 3 Science drive 2, Singapore 117543}
\author{Colin Teo}
\affiliation{Centre for Quantum Technologies, National University of Singapore, 3 Science drive 2, Singapore 117543}
\author{Valerio Scarani}
\affiliation{Centre for Quantum Technologies, National University of Singapore, 3 Science drive 2, Singapore 117543}
\affiliation{Department of Physics, National University of Singapore, 2 Science Drive 3, Singapore 117542}



\begin{abstract}
We study the principle of information-causality in the presence of extremal no-signalling correlations on a tripartite scenario. We prove that all, except one, of the nonlocal correlations lead to violation of information-causality. The remaining non-quantum correlation is shown to satisfy any bipartite physical principle.
\end{abstract}

\maketitle


\section{Introduction}

Quantum theory was developed as a set of mathematical rules from which predictions on physical phenomena could successfully be obtained. Naturally, much effort has been put towards finding intuitive physical principles on which to lay the foundations of quantum theory. In the case of quantum correlations, several such principles were already proposed. For instance, it has been proven that physical correlations satisfy the no-signalling principle \cite{pr} (instantaneous transmission of information between two locations is impossible) and macroscopic locality \cite{miguel} (a macroscopic coarse-graining of quantum correlations can be explained classically). However, these principles are not exclusive of quantum theory: several other theories satisfy them but still allow for non-quantum correlations.

Is there a physical principle that completely identifies the set of quantum correlations?  \emph{Information-causality} (IC) has been proposed as a solution to this question \cite{ic}. The principle of IC imposes a limit on the amount of information-gain an observer (Bob) can reach when another observer (Alice), in a different location, sends him some amount of information. If $m$ bits are sent,  Bob cannot learn more than $m$ bits from Alice's system. This principle is satisfied in the presence of any classical and quantum correlations, but it is not proven to exclude all possible post-quantum correlations \cite{allcock,css}.  Even in the simplest scenario, where correlations are established by two distant observers measuring a pair of two-valued observables on their physical systems, IC might not be sufficient to rule out all non-quantum correlations. So far, it has proven strong enough to exclude a significant portion of supra-quantum theories: for instance, any correlation stronger than the strongest quantum correlation violates IC. More precisely, IC is violated by all correlations violating the Clauser-Horn-Shimony-Holt (CHSH) Bell inequality \cite{chsh} by a value superior to that allowed by quantum theory \cite{ic}. 

Up to now, the search for physical principles defining quantum correlations focused essentially in the bipartite scenario. However, recent results show that the structure of multipartite correlations is much richer than its bipartite correspondent \cite{pbs}. For instance, nontrivial nonlocal games without quantum advantage over classical theory were found \cite{gyni} and the generalization of Gleason's theorem \cite{gleason} for the multipartite case is known to be problematic \cite{gleason3}. It is thus pertinent to ask if information-causality is a good principle in a multipartite scenario.

In the present paper, we apply the principle of IC to correlations arising in the simplest multipartite scenario. It consists of three observers (Alice, Bob and Charlie), each performing two spacelike separated local measurements $(x, y, z = 0, 1)$, with two possible outcomes $(a,b, c=0,1)$, on their local systems. The obtained correlations between the parties' outcomes, conditioned to the choice of observables, is described by the joint probability distribution $P(abc|xyz)$.  Here we assume that the no-signalling principle holds and consider the set of no-signalling extremal correlations (those from which all other no-signalling correlations can be obtained by statistical mixing). We then test the principle of information-causality in the presence of these  correlations. We conclude that information-causality rules out all extremal no-signalling nonlocal tripartite correlations, except one. Using the criterion derived in Ref.~\cite{gallego}, and similarly to an example pointed out there, we show that this singular correlation cannot be ruled out by any bipartite principle.

This paper is organized as follows. In Sec. II we introduce our working tools: we briefly overview the properties of the sets of tripartite probability distributions and present sufficient conditions for the violation of IC principle. In Sec. III, we start by explaining our method. We show how we derive bipartite correlations, on which we can test IC, from tripartite correlations. For that we introduce simple versions of a general processing called \emph{wiring}. We then present our results, namely we show how we are able to exclude all extremal nonlocal tripartite correlations using information-causality, except one. In Sec. IV we analyze this correlation in detail and prove that it satisfies any bipartite criterion. Finally, in Sec. V we discuss the perspectives arising from the present results, specially on the need for a genuine multipartite physical principle.

\section{Review of the Tools}

\subsection{No-signalling, Quantum and Local Tripartite Probability Distributions}

The set of probability distributions for tripartite systems, with binary input and output for each party, is a convex set in a 26-dimensional space, defined by the conditions of positivity ($P(abc|xyz)\geq0$) and normalization ($\sum_{abc}P(abc|xyz)=1, \forall x,y,z$). The set of no-signalling distributions is obtained by imposing that the choice of measurement by one of the parties cannot affect the distribution of the remaining, that is,
\begin{equation}
P(ab|xyz)=\sum_c P(abc|xyz)=P(ab|xy), \forall a,b,x,y,z\,,
\end{equation}
and any permutation of the parties. This defines the \textit{no-signalling polytope}, which has been completely characterized in Ref.\cite{pbs}.  Its complexity is appalling, especially taking into account that it considers just the simplest non-trivial tripartite scenario. Indeed, the no-signalling polytope has 53856 extremal points, belonging to 46 different classes: 45 of them comprise nonlocal points, while the remaining class contains all deterministic local points. These consist of the extreme points of the \emph{local polytope}, i.e. the set of correlations obtained by a local (classical) model
\begin{equation}
P(abc|xyz)=\sum_\lambda p_\lambda P(a|x,\lambda)P(b|y,\lambda)P(c|z,\lambda)\,,
\end{equation}
where $\lambda$ is some random variable, with distribution $p_\lambda$, shared by the parties. Curiously, the local polytope is defined by exactly 46 classes of facets, forming one class of trivial constraints and 45 inequivalent Bell-type inequalities \cite{sliwa}. Unfortunately there is no obvious correspondence between local facets and extremal no-signalling points. 

The set of quantum correlations is the convex body defined by Born's law, $P(abc|xyz)=\tr(\rho M_a^x\otimes M_b^y\otimes M_c^z)$, where $\rho$ is a quantum state and $M$ define quantum measurements. The \emph{quantum set} is clearly not a polytope and little is known about its boundaries.  A convergent hierarchy of semi-definite programs is known \cite{npa}, but there is no guarantee that the quantum set is reached after a finite number of steps. Also, no analytical bounds have been derived and the techniques used in the bipartite case \cite{tzh} would probably not be very tight. Fortunately, for the nonlocal correlations we analyze here, the problem turns out to be simpler and we will always be able to decide on their quantumness.

Our goal is then to determine the ability of  IC principle to exclude post-quantum tripartite correlations, in particular the set of extremal points of the no-signalling polytope \footnote{It is enough to discuss the violation of IC for one representative of each class, since all correlations in the same class are equivalent under relabeling of parties, inputs or outputs.}. Following the classification from Ref.~\cite{pbs}, class 1 corresponds to local deterministic probability distributions, which arise in classical systems and necessarily satisfy IC. We are then interested on the other 45 nonlocal classes, which all turn out to be non-quantum. We will show that 44 of them violate IC (notice that the violation of IC by boxes of class 46, and their natural multipartite generalization, has been studied in detail in Ref.~\cite{china}). The remaining class (class $\sharp 4$) accounts for 126 extremal points and will be analyzed in more detail in Sec. \ref{point4}. There we show that these correlations not only satisfy IC, but any other bipartite information principle aimed at defining the set of quantum correlations.

\subsection{Sufficient Criteria to Violate the IC principle}\label{suff criteria}

The principle of information-causality states that Bob cannot learn more than $m$ bits from a distant Alice, after she sends him $m$ bits. This principle has been formulated based on a specific communication protocol \cite{ic}. It starts with Alice receiving a string of $n$ random and uncorrelated bits, $(a_1,\ldots, a_n)$, and Bob a number between $b$, between 1 and $n$. The goal is for Bob to guess the value of Alice's bit $a_b$. For that, the parties are allowed to share physical resources and operate locally on these, and Alice can send $m$ classical bits to Bob.
Using this scheme, in Ref.s \cite{ic,allcock} it has been proven that information-causality is violated whenever the parties share correlations that either: 
(i) achieve a larger-than-quantum violation of the CHSH inequality, i.e,
\ba
CHSH &=& E_{00}+E_{01}+E_{10}-E_{11}\,>\,2\sqrt{2}\,,\label{chsh}
\ea
where $E_{xy}=P(a=b|xy)-P(a\neq b|xy)$; or
(ii) violate the quadratic inequality proposed by Uffink \cite{uffink}, i.e.
\ba\label{quadratic}
\big(E_{00}+E_{10}\big)^2\,+\,\big(E_{01}-E_{11}\big)^2&>&4\,.
\ea

\section{Tripartite Correlations that Violate Information-Causality} 

\subsection{Approach}
We have seen that information-causality is a principle formulated for bipartite systems, and the simple criteria for its violation (Eqs.\eqref{chsh}-\eqref{quadratic}) are based on correlations between single input/output bits per party. Therefore our tripartite distributions must  be transformed into appropriate bipartite ones before being tested. Our method is the following. First, we bipartite the system by grouping two of the parties together, from which we obtain bipartitions $A|BC$, $AB|C$ or $B|AC$.  As such, the two-party block receives two bits as inputs and outputs two bits.  We transform these into effective single bits by a process called \textit{wiring} \cite{wiring}, where the parties inside the new partitions perform some form of processing (possibly using communication).  Clearly,  communication is forbidden outside the partitions, in order to guarantee that the nonlocal correlations across the bipartition remain the same as in the original tripartite distribution.
More precisely , for each tripartite box $P(a,b,c|x,y,z)$, a bipartition and a wiring define an effective box $P_{\textrm{eff}}(a',b'|x',y')$, where all the inputs and outputs \footnote{Notice that our outputs $a,b,c$ were written $\hat{a},\hat{b},\hat{c}$ in Ref.~\cite{pbs}, where the letters $a,b,c$ were used to denote outcomes belonging to $\{-1,+1\}$.} are considered to be in $\{0,1\}$. 
The goal is to find, for each $P$, a bipartition and a wiring such that $P_{\textrm{eff}}$ violates IC, according to the sufficient criteria in Sec.~\ref{suff criteria}.

We basically use two simple types of wirings, which we will illustrate with examples ahead. Interestingly, like in any other processing, wirings may result in loss of correlations. However, for the purpose of ruling out these extremal correlations, they will prove powerful enough.

\subsection{Results and Examples}

The results are summarized in Table \ref{icviolation}. Only the local deterministic correlations comprising class $\sharp$1 and nonlocal correlations in class $\sharp$4 are not found to violate IC. For all the others, a bipartition and a wiring can be found such that either (\ref{chsh}) or (\ref{quadratic}), or both, happen. We now illustrate the two distinct types of wiring with examples taken from Table \ref{icviolation}. 

\begin{figure}[htbp]
\begin{center}
\includegraphics[scale=0.2]{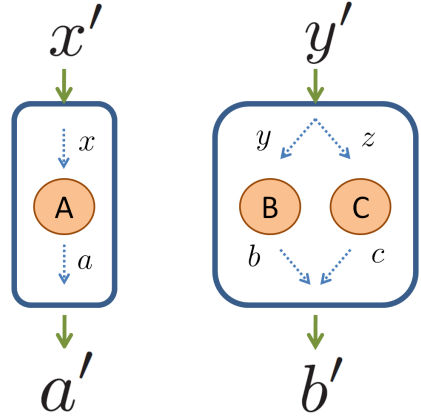}
\end{center}
\caption{\emph{Wiring type I.} Bipartition and wiring that lead to the violation of IC by the extremal points of class $\sharp$44. Here, $A$ has as an input $x=x'$ and outputs $a=a'$. On the other partition however, the input of $C$ is always $z=1$, while the input of B is $y=y'$; the final output is $b'=b\oplus c$.}
\label{fig:wiring1}
\end{figure}

\paragraph{Wiring type I.} Consider a bipartition $A|BC$. In the first type of wiring, the inputs $yz$ of the original tripartite box will only be a function of the effective input $y'$ given to the block BC. Take the wiring on box $\sharp$44, characterized by the relation $a\oplus b\oplus c=x\cdot y \cdot z$, as an example (see Fig.~\ref{fig:wiring1}). The inputs of the tripartite box are defined by $x=x'$, $y=y'$ and $z=1$. Therefore, inside the BC partition, the input $y'$ is given to B and the  input of C is fixed. The outputs are chosen to be $a'=a$ and $b'=b\oplus c$, which realizes $a'\oplus b'=x'\cdot y'$. In other words, using box $\sharp$44 we are actually able to obtain an effective bipartite Popescu-Rohrlich box \cite{pr}, known to violate maximally IC \cite{ic}. Consequently, correlations from box $\sharp44$ are also  forbidden by the principle of information-causality. 

\begin{figure}[htbp]
\begin{center}
\includegraphics[scale=0.2]{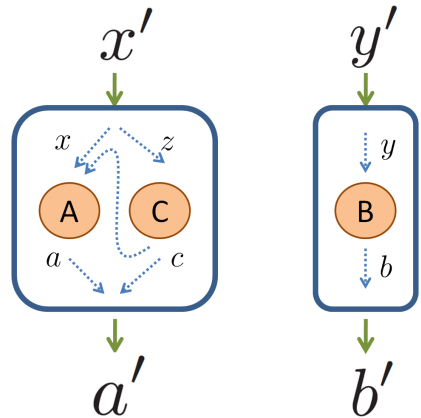}
\end{center}
\caption{\emph{Wiring type II.} Bipartition and wiring that lead to the violation of IC by the extremal points of class $\sharp$3. Here, $B$ has as an input $y=y'$ and outputs $b=b'$. On the other partition, first the input $z=x'$ is used for $C$, then the corresponding output $c$ is used to define the input $x$ of $A$, according to 
$x=x'\cdot(c\oplus 1)$; the final output of this block is $a'=a$.}
\label{fig:wiring2}
\end{figure}

\paragraph{Wiring type II.}  In the second type of wiring, one of the inputs $z$ or $y$ additionally depends on the output of the other party in the block. This necessarily imposes a time order in the use of the tripartite box: if, for example, the input $z$ depends on the output $b$, Charlie can only use his part of the box after receiving the information from Bob's outcome.
A few of the classes of extremal points require this type of wiring to violate IC; its study is slightly more complicated. Take as an example the points of the class $\sharp$3. The explicit form of one of its representatives can be derived from Table 2 in Ref.\cite{pbs}:
\begin{align}
	P(a,b,c|x,y,z) &= \frac{1}{8}\big[1\,+\,(-1)^{a+ b}\,\delta_{x,0}+(-1)^{a+ c}\, \delta_{x,1}\delta_{z,0}+(-1)^{a+ b + c} \,\delta_{x,1}\left(\delta_{y,0}-\delta_{y,1}\right)\delta_{z,1}\big]\,.
\end{align}
The bipartition (now $B|AC$) and wiring are sketched in Fig.~\ref{fig:wiring2}. In Bob's part, $y=y'$ and $b=b'$. In the AC partition, the input $x'$ is first used as the input for $C$, $z=x'$. The output of this box, $c$, will then be used for the input of $A$, according to $x=z\cdot (c\oplus 1)$. The corresponding outcome $a$ is used as final output $a'=a$. 

In order to work out this example, notice that the wiring relation $x=z\cdot (c\oplus 1)$ explicitly reads: if $c=0$, then $x=z$; if $c=1$, then $x=0$ independently of $z$. So:
\begin{align*}
P_{\textrm{eff}}(a',b'|x',y') &= P(a,b,c=0|x=z,y,z)+ P(a,b,c=1|x=0,y,z)\\
&= \frac{1}{8}\big[1\,+\,(-1)^{a+b}\,\delta_{z,0}+(-1)^{a+b} \,\left(\delta_{y,0}-\delta_{y,1}\right)\delta_{z,1}\big]\; +\; \frac{1}{8}\big[1\,+\,(-1)^{a+b}\big]\\
&= \frac{1}{4}\Big[1\,+\,(-1)^{a+b}\,\frac{\delta_{z,0}+\left(\delta_{y,0}-\delta_{y,1}\right)\delta_{z,1}}{2}\Big]\,\equiv\, 
\frac{1}{2}\Big[1\,+\,(-1)^{a+b}\,E_{x'y'}\Big]\,.
\end{align*}
From this last expression, one finds $E_{00}=E_{01}=E_{10}=1$ and $E_{11}=0$, whence $CHSH=3$.

\begin{longtable*}{|c|ccccc|}
  \hline
  \# & Wiring & $a'$ & $b'$ & CHSH & Uffink \\
  \hline
\endfirsthead
\caption{continued.}\\
  \hline
  \# & Wiring & $a'$ & $b'$ & CHSH & Uffink \\
  \hline
\endhead
  \multicolumn{6}{l}{{Continued on the next page\ldots}} \\
\endfoot
  \hline
\endlastfoot
1 & - & - & - & - & - \\
2 & - & $b$ & $c$ & 4 & 8\\
3 & $x=z+zc$ & $a$ & $b$ & 3 & 5 \\
4 & - & - & - & - & - \\
5 & $x=z+zc$ & $a$ & $b$ & 3 & 5 \\
6 & $x=1$ & $a+b$ & $c$ & 4 & 8 \\
7  & $y=1+x$ & $a+b$ & $c$ & 4 & 8 \\
8 & $z=ax$ & $b$ & $c$ & 3 & 5 \\
9 & $y=1+x$ & $a+b$ & $c$ & 4 & 8 \\
10 & $x=0$ & $a+b$ & $c$ & 4 & 8 \\
11 & $z=ax$ & $a+c$ & $b$ & 3 & 5 \\
12 & $z=ax$ & $a+c$ & $b$ & 3 & 5 \\
13 & $y=1+z$ & $a$ & $b+c$ & - & 40/9 \\
14 & $y=1+x$ & $a+b$ & $c$ & 10/3 & 52/9 \\
15 & $x=0$ & $a+b$ & $c$ & 4 & 8 \\
16 & $y=1$ & $a+b$ & $c$ & - & 40/9 \\
17 & $x=0$ & $a+b$ & $c$ & 4 & 8 \\
18 & $z=0$ & $a$ & $b+c$ & 3 & 5 \\
19 & $z=1$ & $a$ & $b+c$ & 3 & 9/2 \\
20 & $z=0$ & $a$ & $b+c$ & 16/5 & 26/5 \\
21 & $z=1$ & $a$ & $b+c$ & 3 & 9/2 \\
22 & $z=1+a+ax$ & $a+c$ & $b$ & - & 40/9 \\
23 & $y=1+x$ & $a+b$ & $c$ & 4 & 8 \\
24 & $x=1$ & $a+b$ & $c$ & 4 & 8 \\
25 & $z=1$ & $a$ & $b+c$ & 10/3 & 52/9 \\
26 & $z=0$ & $a$ & $b+c$ & - & 40/9 \\
27 & $z=1$ & $a$ & $b+c$ & 3 & 5 \\
28 & $x=1$ & $a+b$ & $c$ & 4 & 8 \\
29 & $z=1$ & $a$ & $b+c$ & 10/3 & 52/9 \\
30 & $z=0$ & $a$ & $b+c$ & 18/5 & 114/25 \\
31 & $y=1$ & $a+b$ & $c$ & 14/5 & 4 \\
32 & $z=0$ & $a$ & $b+c$ & 18/5 & 114/25 \\
33 & $y=1$ & $a+b$ & $c$ & 14/5 & 116/25 \\
34 & $z=0$ & $a$ & $b+c$ & 10/3 & 50/9 \\
35 & $z=0$ & $a$ & $b+c$ & 10/3 & 50/9 \\
36 & $z=1$ & $a$ & $b+c$ & 7/2 & 49/8 \\
37 & $z=0$ & $a$ & $b+c$ & 7/2 & 25/4 \\
38 & $z=0$ & $a$ & $b+c$ & 10/3 & 52/9 \\
39 & $z=0$ & $a$ & $b+c$ & 10/3 & 52/9 \\
40 & $z=0$ & $a$ & $b+c$ & 3 & 5 \\
41 & $z=0$ & $a$ & $b+c$ & 3 & 5 \\
42 & $z=0$ & $a$ & $b+c$ & 3 & 5 \\
43 & $z=1$ & $a$ & $b+c$ & 26/7 & 340/49 \\
44 & $z=1$ & $a$ & $b+c$ & 4 & 8 \\
45 & $z=1$ & $a$ & $b+c$ & 4 & 8 \\
46 & $z=1$ & $a$ & $b+c$ & 4 & 8 \\
\hline

\caption{Violation of bipartite IC as detected by the CHSH inequality \eqref{chsh} or the Uffink inequality \eqref{quadratic}. The table follows the conventions of Table 2 in Ref.\cite{pbs}: both the settings $x,y,z$ and the outcomes $a,b,c$, take the values 0 or 1. All the sums are to be taken modulo 2. The bipartitions are implied by the outputs $a',b'$: for instance, if $b'=b+c$, clearly the bipartition must be $A|BC$. Notice that the inequality which is violated may not necessarily be (\ref{chsh}) or (\ref{quadratic}), but one of their equivalent forms under relabeling of the parties and/or the inputs and/or the outputs.}\label{icviolation}
\end{longtable*}


\section{Class $\sharp$4: Extremal No-Signalling Correlations Satisfying any Bipartite Criterion}\label{point4}

We turn now to the detailed study of correlations in class $\sharp4$. We have seen that using our simple techniques we were unable to observe a violation of IC in the presence of these correlations. Since they are supra-quantum (see proof in the Appendix \ref{app}), it is pertinent to ask if different wirings or sharper violation criteria would lead to such a violation. Here, using the results derived in Ref.~\cite{gallego}, we will see this is not the case: non-quantum class $\sharp4$ cannot be excluded by the IC principle. Actually, it cannot be excluded by \emph{any}  bipartite information-theoretical principle aimed at defining the set of quantum (or even local) correlations. But before turning to this, let us describe the points in this class and some of their properties.

\subsection{Description of the Correlations}

The correlations in class $\sharp$4 have a very precise structure. As its representative, we choose the one that is obtained from the version in Table 2 in Ref.\cite{pbs} by changing $y\rightarrow 1-y$. In this formulation, the class $\sharp$4 is defined by the following deterministic correlations
\begin{align}
	a_0\oplus b_1 &= 0, \nonumber \\
	b_0\oplus c_1 &= 0, \nonumber\\
	c_0\oplus a_1 &= 0, \nonumber\\
	a_0\oplus b_0\oplus c_0 &= 0, \nonumber\\
	a_1\oplus b_1\oplus c_1 &= 1, \label{class4}
\end{align}
and random statistics for any other combination. Here, $a_x$ here refers to the output of $A$ when its input is $x$ and similarly for the $B$ and $C$. It is clear from these equations that any cyclic permutations of $(A,B,C)$ will have the same statistics.

Class $\sharp4$  is clearly non-local, because the sum of the first four correlations would imply $a_1\oplus b_1\oplus c_1=0$, which contradicts the last relation. It also cannot be realized within quantum theory , see Appendix \ref{app} for details. Another remarkable property is that these correlations cannot be created by distributing arbitrary many bipartite PR-boxes between the parties \cite{pironioprivate}. What we prove in the following subsection yields as a corollary that the converse is also true: these correlations cannot be used to create a bipartite PR-box.

\subsection{No Violation by Any Bipartite Criterion}

We now show that extremal points in class $\sharp$4 satisfy IC and any other bipartite information principle aimed at singling out quantum correlations. The main idea is to prove that the probability distributions belonging to this class are local for any bipartition, even after any local wirings. A sufficient criterion was provided in Ref.~\cite{gallego}: if $P(a,b,c|x,y,z)$ belongs to the set of \textit{time-ordered bi-local} (TOBL) probability distributions, all possible bipartite distributions derived from $P(a,b,c|x,y,z)$ are local. A probability distribution belongs to TOBL if it can be written as
\begin{align} \label{tobl}
	P(a,b,c|x,y,z) &= \sum_\lambda p_\lambda \; P(a|x,\lambda) \; P_{B}(b|y,\lambda) \;P_{C}(c|b,y,z,\lambda)\,,  \\
	&= \sum_\lambda p_{\lambda} \; P(a|x,\lambda) \; P'_{B}(b|c,y,z,\lambda) \;P'_{C}(c|z,\lambda),\notag
\end{align}
for bipartition $A|BC$, and analogously for $B|AC$ and $C|AB$. Here $p_\lambda$ is the probability distribution over the shared random variable $\lambda$. We see that the model allows signalling from Bob to Charlie (first line) or from Charlie to Bob (second line), but never both at the same time. This accounts for the unknown time-ordered events. Also, Alice is unable to infer the direction of signalling inside the bipartition $BC$. 

We show that correlations in class $\sharp$4 belong to TOBL by constructing a model \eqref{tobl} that generates its statistics. Let the hidden variable $\lambda$ be a vector of two bits $\lambda=(\lambda_0,\lambda_1)$, distributed along the uniform probability distribution $p(\lambda)=\frac{1}{4}$.  To start, consider the splitting $A|BC$. Alice always outputs according to the strategy
\begin{equation}
	a = \lambda_0 \oplus(\lambda_0\oplus \lambda_1)\cdot x\,.
\end{equation}
Inside the partition $BC$, the instructions change according to the direction of signalling.  If Bob receives his input before Charlie, it must be independent of $z$ and $c$, therefore only $B\rightarrow C$ signalling is possible. In this case, the strategy is
\begin{align}
	b &= \lambda_0 \oplus \lambda_1 \oplus \lambda_1\cdot y, \notag\\
	c &= \lambda_1 \oplus (\lambda_0 \oplus y)\cdot z\,.
\end{align}
If Charlie receives his input before Bob ($C\rightarrow B$), they follow the instructions
\begin{align}
	b &= \lambda_0 \oplus (\lambda_1 \oplus z)\cdot (y\oplus 1)\notag\\
        c &= \lambda_1 \oplus (\lambda_0 \oplus 1)\cdot z\,.
\end{align}
Since correlations in class $\sharp4$ are invariant under cyclic permutation of the parties $(A,B,C)$, similar models are valid for the bipartitions $C|AB$ and $B|CA$. We can easily check that this completely specifies a TOBL model \eqref{tobl} that reproduces the correlations of class $\sharp4$ \eqref{class4}. We observe these have local (classical) statistics for any bipartition we consider, even after wirings, so they will always respect any bipartite information-theoretical principle aimed to single out quantum (or even local) correlations. 

A non-extremal correlation with the same property has already been found in another region of the polytope \cite{gallego}, specifically above the \emph{guess-your-neighbor-input} (GYNI) tripartite inequality \cite{gyni},
\be
P(000|000)+P(110|011)+P(011|101)+P(101|110)\leq1\,.
\ee  
Notice that this conclusion extends at least to all points which are convex combinations of those examples and some local deterministic points. Then, the non-quantum correlations that cannot be ruled out with bipartite criteria form sets of non-zero measure.

To finish, we show that our class of extremal points does not violate \textit{any} GYNI inequality, which proves that our example and the one in Ref. \cite{gallego} exhibit intrinsically different kinds of nonlocality. First, we recall that the most general form of tripartite GYNI inequalities \cite{gyni} is
\be\label{gen gyni}
\sum_{x_1,x_2,x_3=0,1} q(x_1,x_2,x_3)P(x_2x_3x_1|x_1x_2x_3)\leq \max_{x_1,x_2,x_3}(q_{x_1,x_2,x_3}+q_{\bar x_1,\bar x_2,\bar x_3})
\ee
where $\sum_{x_1,x_2,x_3}q(x_1,x_2,x_3)=1$ and the upper bars denote negation. 
For the points in class $\sharp$4, the tripartite probabilities are never larger than 1/4: $P(a_1,a_2,a_3|x_1,x_2,x_3)\leq\frac{1}{4}$. So, in the best case, these correlations provide the value 
\be
\frac{1}{4}\big(\sum_{x_1\oplus x_2\oplus x_3=0}(q_{x_1,x_2,x_3}+q_{\bar x_1,\bar x_2,\bar x_3})\big)
\ee
to the left-hand side of Eq.\ref{gen gyni}. Clearly, never exceeds the local bound $\max_{x_1,x_2,x_3}(q_{x_1,x_2,x_3}+q_{\bar x_1,\bar x_2,\bar x_3})$ and nonlocal boxes in class $\sharp$4 are unable to violate GYNI inequalities.

\section{Perspectives and Conclusions}

We have set out to apply the principle of information-causality to multipartite correlations. We took the only existing form to test IC, which involves a bipartite task, and checked its violation on the extremal points of the simplest tripartite no-signalling polytope. IC principle excludes 44 out of 45 classes of non-local extremal points (which consequently are non-quantum); the remaining class, which can also be proved not to be quantum by other means, satisfies this and any other bipartite principle.

IC remains a powerful criterion to rule out no-signalling correlations which cannot be achieved within quantum physics. But it started out with a more ambitious conjecture, namely, as a physical principle which might identify the set of quantum correlations exactly. Generalizations of IC beyond the basic CHSH scenario cannot be said to have brought clarification: instead, we are discovering extremely complex structures. These are not devoid of interest and deserve further studies; certainly, for instance, it will be very instructive to find a meaningful multipartite task that detects the non-quantumness of our example and of the one reported in Ref.~\cite{gallego}. However, ultimately, completely different approaches will probably have to be found, in order to prove the main conjecture.

\section*{Acknowledgments}

We acknowledge illuminating discussions with Antonio Ac\'{\i}n, Nicolas Brunner, Rodrigo Gallego, Miguel Navascu\'es and Lars W\"urflinger. We thank one of the anonymous referees for pointing out the elegant argument used in the proof in Appendix \ref{app}. This work was supported by the National Research Foundation and the Ministry of Education, Singapore.

\begin{appendix}
\section{Class $\sharp$4 is non-quantum.}\label{app}

Here we show that probability distributions of class $\sharp$4 cannot be obtained within quantum physics. For that, we will see that they violate an inequality satisfied by any local and quantum correlations. This inequality reads as follows,
\begin{align}\label{quantuminequality}
	k=\frac{15}{2} + \frac{1}{2}\expect{ A_1B_1C_1} - 2 (\expect{ A_0B_0C_0}+\expect{A_0B_1} + \expect{B_0C_1} + \expect{A_1C_0} )\geq0
\end{align}
where $A_i$, $B_i$ and $C_i$ are local observables with possible values $\pm1$.
To prove that it holds true for any quantum observables, we define the operator $K$ as
 \begin{align}
	K \equiv \left( \frac{\alpha\beta+\gamma\delta}{2} -1 \right)\left( \frac{\alpha\beta+\gamma\delta}{2} -1 \right)^\dagger+ 2 \left( \frac{\alpha+\beta}{2} -1 \right)^2 + 2 \left( \frac{\gamma+\delta}{2} -1 \right)^2
\end{align}
where $1$ is the identity operator while $\alpha,\beta,\gamma,\delta$ are any arbitrary operators. One can easily check by inspection that $K$ is positive semidefinite and $\expect{K}\geq0$. Now, performing the substitution
\begin{align}
	\alpha &=A_1C_0, \nonumber\\
	\beta &=A_0B_1, \nonumber\\
	\gamma &=A_0B_0C_0, \\
	\delta &=B_0C_1\nonumber\,,
\end{align}
and some simple algebra, we observe that $\expect{K}$ defines exactly the expression \eqref{quantuminequality}. Hence,  inequality \eqref{quantuminequality} is valid for any theory where local observables are defined by operators on a Hilbert space. In particular, quantum theory.

Consider now the correlations that define class $\sharp$4. In the case where local observables have possible outcomes $\pm1$, relations \eqref{class4} can be transformed into
\begin{align}
\expect{A_0B_1}&=1\nonumber\\
\expect{C_0A_1}&=1\nonumber\\
\expect{B_0C_1}&=1 \label{class4ph}\\ 
\expect{A_0B_0C_0}&=1\nonumber\\
\expect{A_1B_1C_1}&=-1\nonumber\,.
\end{align}
Simple substitution in \eqref{quantuminequality} gives $\langle K\rangle = -1$, from which we conclude that class $\sharp$4 cannot be quantum.

\end{appendix}

\end{document}